\documentclass[%
 reprint,
 amsmath,amssymb,
 aps,
]{revtex4-2}

\usepackage{graphicx}
\usepackage{dcolumn}
\usepackage{bm}
\usepackage{mathtools}
\usepackage{xcolor}

\newcommand\physrep{Phys.~Rep.}      
\newcommand\ssr{Space Sci. Rev.}     
\newcommand\apss{Ap\&SS}             
\newcommand\jcap{J.~Cosmology Astropart. Phys.} 

\begin{document}

\preprint{APS/123-QED}

\title{Relative efficiency of three mechanisms of vector fields growth in a random media}

\author{Illarionov E.A.}
\affiliation{
 Department of Mechanics and Mathematics, Moscow State University, Moscow, Russia
\\
Moscow Center of Fundamental and Applied Mathematics, Moscow, Russia}
\author{Sokoloff D.D.}
  \affiliation{Department of Physics, Moscow State University, Moscow, Russia \\
  Moscow Center of Fundamental and Applied Mathematics, Moscow, Russia}

\date{\today}

\begin{abstract}
We consider a model of a random media with fixed and finite memory time with abrupt losses of memory (renovation model).
Within the memory intervals we can observe either amplification or oscillation of the vector field in a given particle.
The cumulative effect of amplifications in many subsequent intervals leads
to amplification of the mean field and mean energy. Similarly, the cumulative effect
of intermittent amplifications or oscillations also leads to amplification
of the mean field and mean energy, however, at a lower rate.  Finally, the random oscillations alone 
can resonate and yield the growth of the mean field and energy. 
These are the three mechanisms that we investigate and
compute analytically and numerically the growth rates based on 
the Jacobi equation with the random curvature parameter. 
\end{abstract}

\maketitle

\section{Introduction}

Instabilities in random media often arise in various branches of physics, in particular in the context of 
astrophysical magnetohydrodynamics (MHD). To be specific, one can discuss instabilities in random media in the
context of, say,  magnetic field self-excitation in celestial bodies, namely the dynamo theory (e.g. \cite{BS05, S04}).
Convective motions of electrically conductive media  
are believed to generate a cyclically variable solar magnetic field.
The process known as the solar dynamo combines deterministic (e.g., the almost periodic variability of the monthly sunspot number, known as the solar cycle) and random (e.g., the apparently random variation of the cycle amplitude from one cycle to another) features. 
The process is widely investigated on the basis of astronomical observations and
specific astrophysical models, confirmed by numerical simulations, are proposed to explain the details.
The basic concept of dynamo is confirmed by laboratory experiments (among many others see, e.g., \cite{Betal12, Retal13}).      
At the same time, abilities of laboratory and numerical
experiments are fundamentally limited regarding the context
of astrophysical (and even laboratory) phenomena. Thus the analytical results and
estimates, even taking into account the simplified and idealized
problem statement, remain and will remain actual.

As early as 1964,  Zeldovich \citep{Z64} proposed a simple model that mimics a lot of basic properties of instabilities in random media and is simple enough for analytical investigation.  
The model qualitatively describes how the intermittent structure of the universe
(i.e. rare mass concentrations represented in the model by random curvature fluctuations) can lead to the systematic
divergence of the two initially close trajectories of light. In the series of
later publications \cite{Letal03,Letal05} this model was formulated in the precise mathematical
terms that enable the detailed analytical and numerical investigation.
More detailed, the problem was formulated in terms of the Jacobi equation
with the random curvature parameter. Various assumptions on the 
distribution of the curvature parameter and its memory time lead
to a wide bulk of practical models. The practical benefit of these models
is that many phenomena observed in such simple models are then also
reproduced in much complex models, in particular, in MHD and dynamo models (e.g., \cite{Ietal22}). 
The point is that a complex evolutionary MHD problem can be reduced within the framework of the Lagrangian approach
to a system of ordinary differential equations. If we are interested in instability, the equations should be linear.
Random pulsations associated with convection or turbulence introduce noise into the problem. Meanwhile,
the probability theory shows that the properties of linear differential equations with random coefficients
are quite general. 

In this work, we focus on quantitative estimation of various mechanisms that
lead to amplification of the mean vector field and energy in random media.
To be specific, we consider the question in the framework of the Jacobi equation
with the random curvature parameter. We consider the renovation model,
in which the curvature parameter remains constant in the intervals of length $\Delta$
and constant values are sampled independently and with the same distribution for each
interval. For the negative curvature two initially close geodetic lines diverge
exponentially fast and thus the solution of the Jacobi equation (known as the Jacobi field)
also growth exponentially. For the positive curvature the Jacobi field oscillates.
The main result of  \cite{Z64}  demonstrates that
intervals with intermittent positive and negative curvature still yield exponential growth.
Recently, in \cite{SI15, SCI21} the growth rates of the Jacobi field and its higher statistical
moments were estimated analytically and numerically.

In this work, we apply the methods proposed in \cite{SI15} and \cite{SCI21} to the three
scenarios of the field growth. The first one assumes that the
curvature parameter is distributed in the negative domain entirely. In the second case the 
curvature parameter has a symmetric distribution. The third case assumes that the
curvature parameter is distributed in the positive domain entirely.
In the latter case, random oscillations resonate and yield
moderate but still exponential growth of the mean Jacobi field and its higher statistical moments.
The main result of this work is a comparison of the relative efficiency of all three
mechanisms described above.

Note that the paper \cite{Z64} is interesting in various aspects, in particular as an early approach to the problem of gravitation lensing (e.g., \cite{KK18}), however, that aspects do not belong to the problem discussed below.

\section{The Jacobi equation}

Below we recall the Jacobi equation and idea of the renovation model.
The Jacobi equation considered for $x\in[0, +\infty)$ reads
\begin{equation}
    y'' + K(x)y = 0 \, .
\label{eq:jac}
\end{equation}
We assume that $K(x)$ is a random process constant at intervals $\Delta_n = [n\Delta, (n+1)\Delta)$ referred to as renovation intervals. The parameter $\Delta$ can be considered as the memory scale. Constant values $K(x) = K_n$ at these intervals are independent random variables with the same distribution. 

Note that canonically the Jacobi equation describes the divergence of two geodesic lines propagating from the same point of two-dimensional Riemannian space.
In these term, the parameter $K$ is related to the Gaussian curvature along the geodesic line (see, e.g., \cite{Klingenberg1995}).

The solution $y(x)$ of the Jacobi equation is known as the Jacobi field. 
The mean growth rate of the Jacobi field is estimated by
the Lyapunov exponent
\begin{equation}
    \lambda =  \lim\limits_{n\to\infty} \frac{1}{n\Delta}\langle \ln |y(n\Delta)| \rangle\, .
\label{eq:lyap}
\end{equation}

Additionally, growth rates of statistical moment of the order $p$ of the  Jacobi field are defined as follows:
\begin{equation}
    \gamma_p =  \lim\limits_{n\to\infty} \frac{1}{2pn\Delta} \ln \langle |y(n\Delta)|^p \rangle\, .
\label{eq:moments}
\end{equation}
Below we will consider only the second moment ($p=2$), which can be interpreted as the mean energy, however, the methods discussed allow extension on the higher moments. 

Introduce auxiliary variables $z_1 = y$, $z_2 = y'\Delta$ to obtain from Eq.~(\ref{eq:jac}) a system of the first order equations: 
\begin{equation}
{ d \over {dx}}\begin{pmatrix}z_1, z_2\end{pmatrix}  = \begin{pmatrix}z_1, z_2\end{pmatrix} \begin{pmatrix}
0 &  -K\Delta \\
1/\Delta &  0 
\end{pmatrix} \, .
\label{eq:jac_system}
\end{equation}
In matrix notation we rewrite (\ref{eq:jac_system}) as
\begin{equation}
{\bf z}'  = {\bf z}B \, .
\label{eq:jac_Z}
\end{equation}
The solution of this system is
\begin{equation}
{\bf z}(n\Delta)  = {\bf z}_0\exp(B_1\Delta)\exp(B_2\Delta)...\exp(B_n\Delta) \, ,
\label{eq:sol}
\end{equation}
where $B_i$ are realizations of the random matrix $B$ at the $i$-th renovation interval. 
Denoting $k = K\Delta^2$, the matrix exponential $\hat B = \exp(B\Delta)$ can be expressed explicitly
\begin{equation}
\hat B = \begin{pmatrix} 
 \cos \sqrt{k}   &  -\sqrt{k} \sin\sqrt{k}  \\
 \frac{\sin\sqrt{k}}{\sqrt{k}}  &  \cos\sqrt{k}
 \end{pmatrix} \, , 
\label{eq:mexp}
\end{equation}
that simplifies investigation of the solutions. Note while $\sqrt k$ can be complex,
the matrix $\hat B$  remains real.

\begin{figure}[h]
\centering
\includegraphics[width=0.48\textwidth]{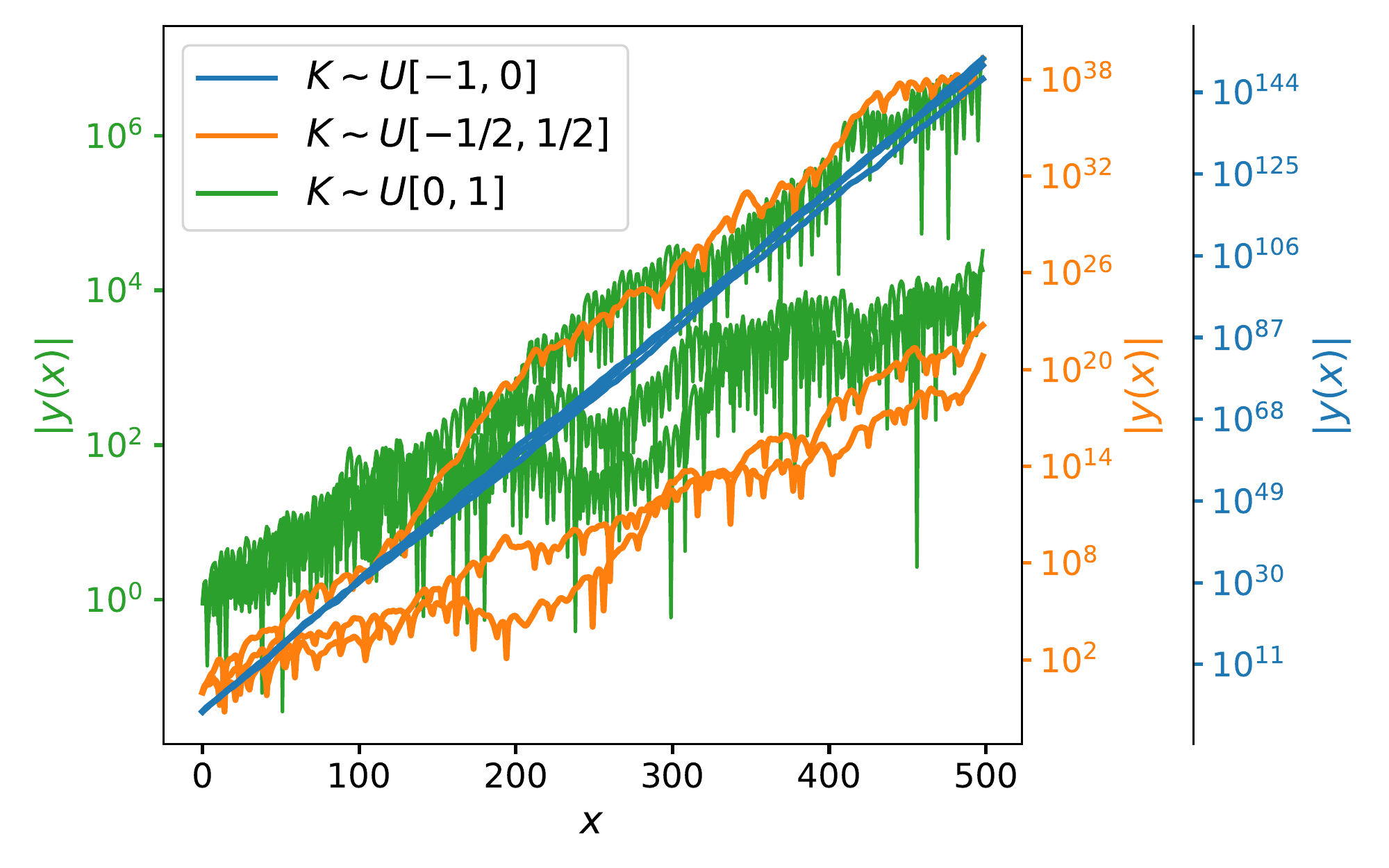}
\caption{Absolute values of sample realizations of the Jacobi field for different distributions of $K$.
The colors correspond to different distributions of $K$. The absolute values of three sample realizations are
shown for each distribution. Note the logarithmic scale for the $y$ axes and
that each distribution is shown with a separate $y$ axis
(the colors of the axes correspond to the colors of the plots).
Separate peaks downwards correspond to moments when the Jacobi field changes sign and
the absolute value of the Jacobi field approaches zero.
}
\label{fig:sample}
\end{figure}

For illustration of the results obtained later in this paper, we will
consider three particular distributions of $K$ corresponding to the 
three scenarios under discussion, namely, we will consider $K$
with a uniform distribution in the negative domain $[-1, 0]$,
in the symmetric domain $[-1/2, 1/2]$, and in the positive domain $[0, 1]$.
In all cases, $K$ has the same standard deviation.
Fig.~\ref{fig:sample}  
shows sample realizations of the Jacobi field $|y(x)|$ obtained from (\ref{eq:sol})
for each case of the distribution of $K$. The realizations were obtained in 
the master's project of our student E.~Makarenko (Moscow State University). 
It can be noted that all realizations
are growing and the growth rate is specific for each probability distribution. The aim of the next parts of the paper is to quantify and compare the 
growth rates.

\section{Invariant measure and the Lyapunov exponent}

The mathematical theory on a product of random matrices developed by
Furstenberg and Kesten in \cite{FK60, F63} allows reduction of the Lyapunov
exponent from the limit expression (\ref{eq:lyap}) to the integration
over a particular invariant measure (also known as the stationary distribution). 
In a simplified notation proposed in \cite{tutubalin1977}, it was shown that the sequence of the matrix exponentials $\hat B_i$ generates an ergodic Markovian chain on the unit sphere $S^1$ with identified
diametrically opposite points. The ergodicity ensures that any initial distribution on this space converges to the unique stationary distribution (invariant measure). 

In \cite{SI15} it was proposed to obtain the transition density $p(\varphi, \psi)$ of the Markovian chain numerically and then to obtain the solution $\pi(\psi)$ of the stationary distribution equation
\begin{equation}
    \pi(\psi) = \int \pi(\varphi) p(\varphi, \psi) {\rm d}\varphi \, .
\end{equation}
Here $\varphi$ and $\psi$ are in $(-\pi/2, \pi/2]$ and specify the coordinates on the
the unit sphere $S^1$ with identified diametrically opposite points.

To derive the transition density, one can demonstrate that the variables $\varphi$ and $\psi$ are related by equation
\begin{equation}
    \tan\psi = \frac{\tan\varphi-\xi\tan\sqrt{k}}{1+\tan\varphi\tan\sqrt{k}/\sqrt{k}} \, ,
\label{eq:trans}
\end{equation}
where $k=K\Delta^2$. Thus, given a distribution of $K$ one can compute the transition density $p(\varphi, \psi)$. 

\begin{figure}[h]
\centering
\includegraphics[width=0.48\textwidth]{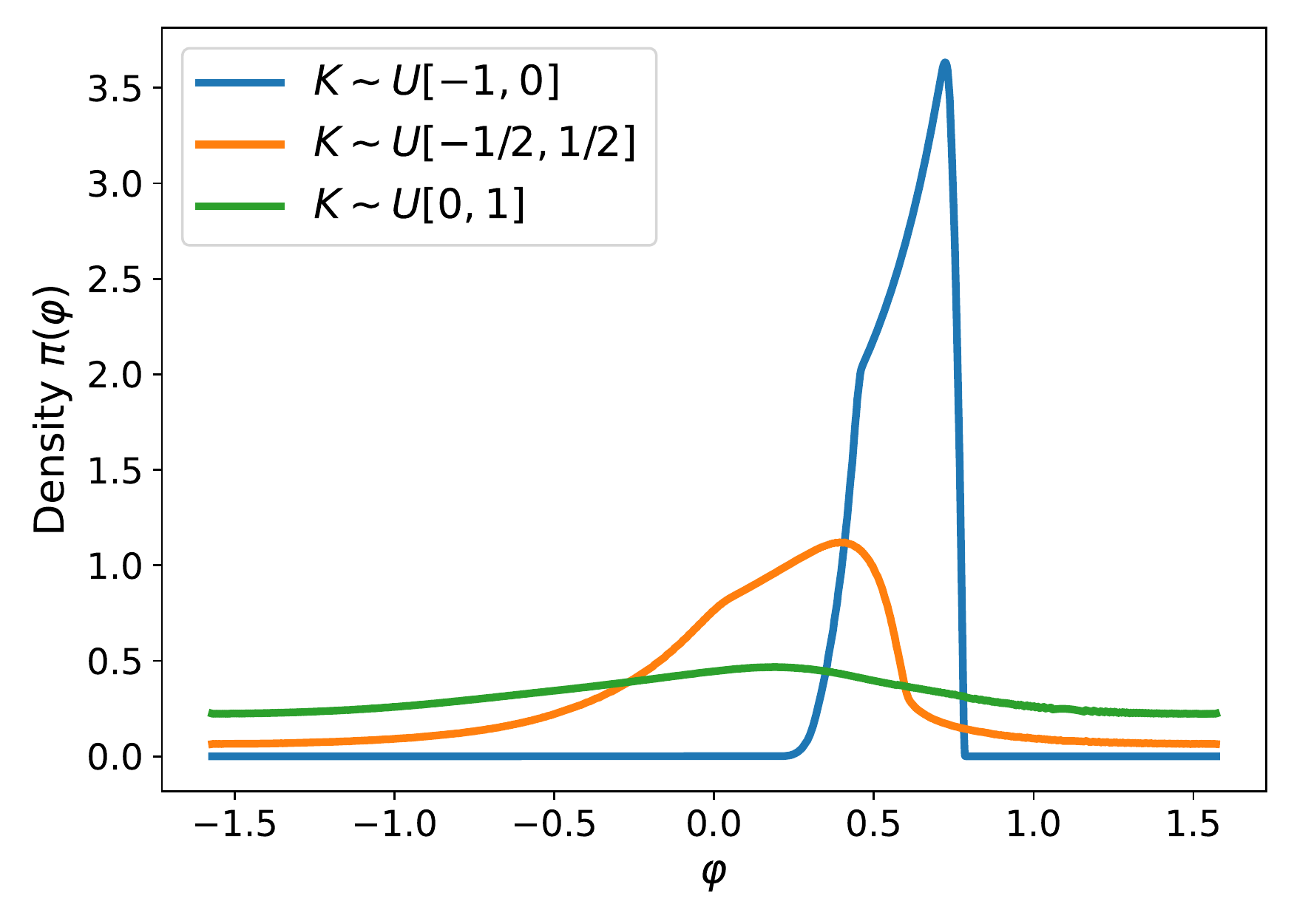}
\caption{Stationary distribution $\pi(\varphi)$ for various distributions of $K$.}
\label{fig:inv}
\end{figure}

In general, the transition density can not be expressed analytically and should be found numerically. In fact, however, the numerical approach  becomes tricky due to various irregularities of the Eq.~(\ref{eq:trans}). We skip the technical details of this process, and present in Fig.~\ref{fig:inv} the final numerical results for the stationary distribution.

Given the stationary distribution $\pi(\varphi)$ and the distribution $p(K)$ of the curvature parameter, we compute the Lyapunov exponent by numerical approximation of the integral
\begin{equation}
    \lambda = \int \log\|{\bf w}{B(K)}\|\pi(\varphi)p(K){\rm d}\varphi {\rm d}K\, ,
\end{equation}
where ${\bf w}$ is the unit vector $(\cos\varphi, \sin\varphi)$.

In Table~\ref{tab:lyap} we summarize the Lyapunov exponents obtained for uniform
distributions of $K$ given that $\Delta=1$. In agreement with Fig.~\ref{fig:sample}
we observe in all cases a positive mean growth rate (the Lyapunov exponent) of the Jacobi field.
For positive $K$ this is the manifestation of the resonance of oscillating solutions,
while for negative and alternating $K$ this indicates that the exponential growth
on the intervals, where $K$ is negative, gives the main contribution to the mean growth rate. Quantitatively,
we observe that the growth rate due to resonance is an order of magnitude
weaker compared to the alternative scenarios.
\begin{table}[]
    \centering
    \begin{tabular}{l|c}
        $K$ & $\lambda$ \\
        \hline
     $U[-1, 0]$ & 0.693\\
     $U[-1/2, 1/2]$ & 0.133\\
     $U[0, 1]$ & 0.018
    \end{tabular}
    \caption{Lyapunov exponents of the Jacobi field for various distributions of $K$.}
\label{tab:lyap}
\end{table}

\section{Growth rate of the mean field}

The positive values of the Lyapunov exponent imply that the 
absolute values of individual realizations of the Jacobi field are expected to grow
exponentially. At the same time, the mean value of the Jacobi field can 
be vanishing as we demonstrate below.

Consider the averaging of Eq.~(\ref{eq:sol}):
\begin{equation}
    \langle{\bf z}\rangle(n\Delta) = {\bf z}_0 \langle\hat B\rangle^n \,  .
\end{equation}
This shows that the mean field $\langle{\bf z}\rangle$ grows as the leading
eigenvalue of the matrix $\langle\hat B\rangle$. More precisely, the growth rate
$\gamma$ of the mean field $\langle{\bf z}\rangle$ is obtained as
\begin{equation}
    \gamma = \frac{1}{2\Delta}\ln|\lambda_{(1)}| \, ,
\label{eq:gamma_mean}
\end{equation}
where $\lambda_{(1)}$ is the leading eigenvalue of the matrix $\langle\hat B\rangle$.
Note that the growth rate $\gamma$ of the mean Jacobi field should not be confused with the
growth rate $\gamma_1$ defined in (\ref{eq:moments}). The difference is that $\gamma_1$
refers to the growth of the mean absolute value of the Jacobi field, while $\gamma$
refers to the growth of the absolute value of the mean Jacobi field.

For small $\Delta$ the matrix $\langle\hat B\rangle$ and then $\lambda_{(1)}$
and $\gamma$ can be approximated in terms of the first several
statistical moments of the random variable $K$, i.e. in terms of $\mu=\langle K \rangle$
and $\sigma^2=var K$. We consider the 
cases when $K$ is distributed (1) in the negative domain, (2) in the domain symmetric with respect to zero, and (3) in the positive domain. 

The first case ($K$ is negative) yields
\begin{equation}
    \gamma = \frac{1}{2}\sqrt{-\mu} +
    \frac{1}{24}\frac{\sigma^2}{\sqrt {-\mu}}\Delta^2
     + O(\Delta^{3})\, .
\label{gamma_neg}
\end{equation}
Note that $\gamma$ does not vanish for $\Delta\to0$ but tends to some positive value.
We will address this fact in the next section.

For the symmetric-distributed $K$ we obtain
\begin{equation}
    \gamma = \frac{\sqrt 6}{12}\sigma\Delta 
    + O(\Delta^{3})\, ,
\label{gamma_sim}
\end{equation}
and $\gamma$ vanishes for $\Delta\to0$.

Finally, for the positive $K$ we obtain
\begin{equation}
    \gamma = -\frac{1}{48}\sigma^2\Delta^3 + O(\Delta^{4})\, .
\label{gamma_pos}
\end{equation}
Here we note that for small $\Delta$ the growth rate $\gamma$ becomes negative.

\begin{figure}[h]
\centering
\includegraphics[width=0.48\textwidth]{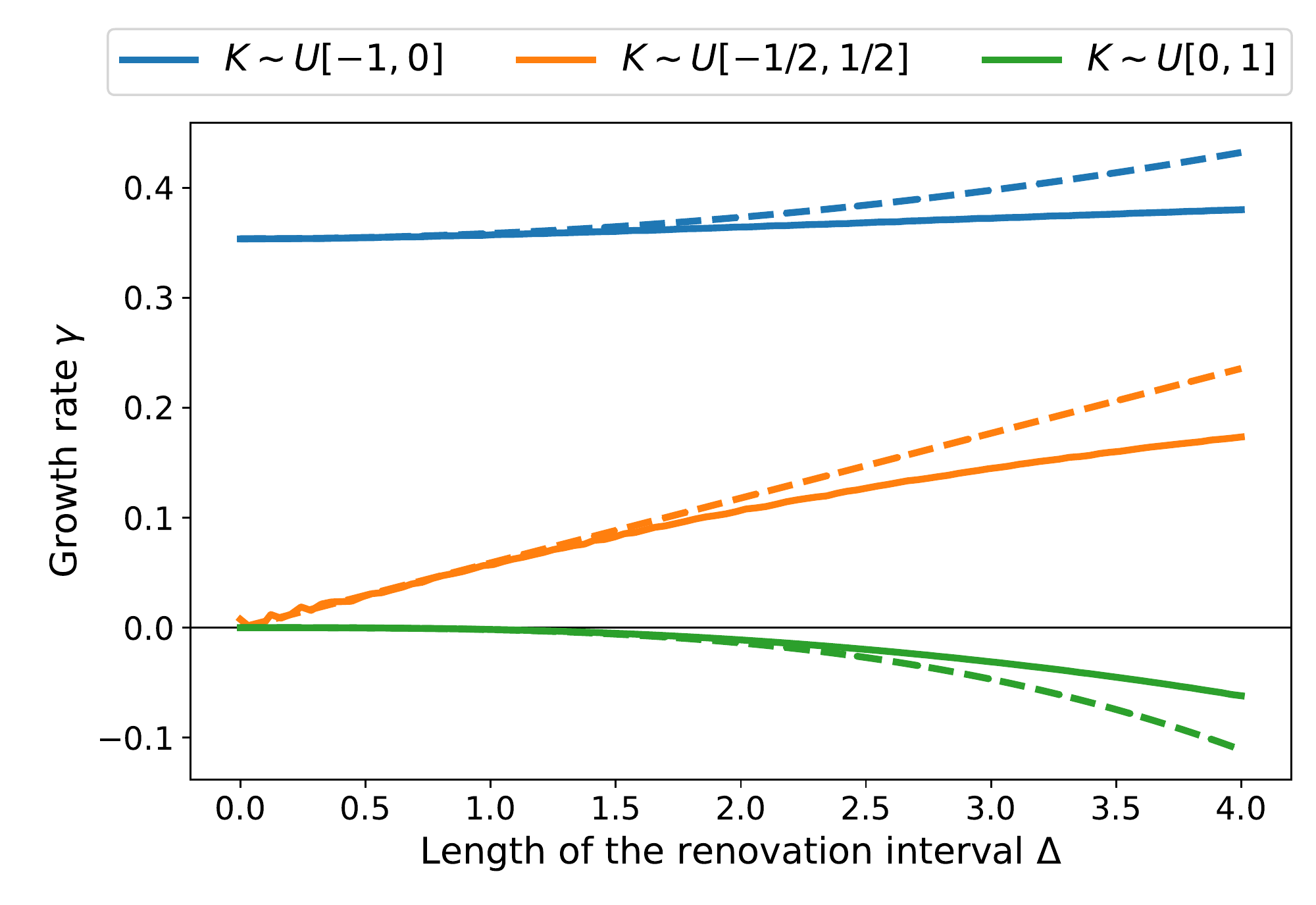}
\caption{Growth rate $\gamma$ of the mean Jacobi field for different distributions of $K$.
Solid lines show numerical estimates, dashed lines show analytical approximations. }
\label{fig:gamma1}
\end{figure}

For illustration, consider uniform distributions for the parameter $K$,
similar to those we considered for numerical estimation of the Lyapunov exponent.
In all cases $\sigma^2=1/12$, but $\mu=-1/2$ for $K\sim U[-1,0]$,
$\mu=0$ for $K\sim U[-1/2,1/2]$, and $\mu=1/2$ for $K\sim U[0, 1]$.
Fig.~\ref{fig:gamma1} shows the analytical approximations of $\gamma$
in comparison with its numerical estimates. For the numerical estimates
we simulated $10^6$ realization of random matrices $\hat B$, 
computed the leading eigenvalue of the sample averaged matrix, and then
derived $\gamma$. We observe that the analytical 
approximations (\ref{gamma_neg})--(\ref{gamma_pos}) are in agreement with 
the numerical simulation at least for small $\Delta$. Of course, one can improve the agreement
by adding terms of the higher orders in (\ref{gamma_neg})--(\ref{gamma_pos}),
however, these terms will include higher statistical moments of the random variable~$K$. 

\section{Second-order statistical moment}

Following \cite{SCI21}, the growth rate of $\langle \|{\bf z} \| ^2 \rangle$
can be associated with the growth rate of matrix product. To show this, 
introduce a vector ${\bf q}$ of pairwise products of the components of vector ${\bf z}$, i.e.
\begin{equation}
    {\bf q} = (z_1z_1, z_1z_2, z_2z_1, z_2z_2) \, .
\end{equation}
Taking the derivative of ${\bf q}$ and using the Jacobi equation in the form of (\ref{eq:jac_Z}),
one obtains the equation
\begin{equation}
    {\bf q}' = {\bf q} A \, .
\end{equation}
Similar to (\ref{eq:sol}), the solution $q(n\Delta)$ is represented by the product of
independent and identically distributed matrix exponentials $\exp(A_i\Delta)$.
It follows that 
\begin{equation}
    \langle{\bf q}\rangle(n\Delta) = q_0 \langle \exp (A\Delta) \rangle^n \, 
\end{equation}
and the growth rate of the mean vector $\langle{\bf q}\rangle$ is given by the leading eigenvalue of the mean matrix exponential $\langle \exp (A\Delta) \rangle$. But
the first component of the vector $\langle{\bf q}\rangle$ is $\langle z_1^2 \rangle=\langle y^2 \rangle$ and thus the leading eigenvalue also defines the growth rate of $\langle y^2 \rangle$.

The matrix exponential $\hat A = \exp (A\Delta)$ can be found explicitly (see \cite{SCI21}) and reads
\begin{equation}
\hat A = \begin{pmatrix} 
 \cos^2 \sqrt{k}   &    \frac{\sin 2\sqrt{k}}{2 \sqrt{k}}   &   \frac{\sin 2 \sqrt{k}}{2 \sqrt{k}}  &  \frac{\sin^2\sqrt{k}}{k} \\
 
 \frac{-\sqrt{k}\sin2\sqrt{k}}{2}   &   \cos^2\sqrt{k}   &   -\sin^2 \sqrt{k} &  \frac{\sin 2 \sqrt{k}}{2 \sqrt{k}} \\
 
 \frac{-\sqrt{k}\sin2\sqrt{k}}{2}   &   -\sin^2 \sqrt{k}   &   \cos^2\sqrt{k}  &  \frac{\sin 2 \sqrt{k}}{2 \sqrt{k}}\\
 
  k \sin^2\sqrt{k}    &    \frac{-\sqrt{k}\sin2\sqrt{k}}{2}   &   \frac{-\sqrt{k}\sin2\sqrt{k}}{2}   &  \cos^2\sqrt{k}
 \end{pmatrix} \, , 
\label{posK}
\end{equation}
where $k = K\Delta^2$ similar to (\ref{eq:mexp}). For small $\Delta$ the mean matrix 
exponential  $\langle \hat A \rangle$ and its leading eigenvalue $\lambda_{(1)}$ can
be approximated in terms of the statistical
moments of the random variable $K$. Then taking into account the relation between the
leading eigenvalue $\lambda_{(1)}$ and the growth rate $\gamma_2$ 
\begin{equation}
    \gamma_2 = \frac{1}{4\Delta}\ln|\lambda_{(1)}| \, ,
\end{equation}
we obtain the following estimates.

For $K$ distributed in the negative domain (note that $\mu < 0$ in this case) we obtain
\begin{equation}
    \gamma_2 =\frac{1}{2}\sqrt{-\mu} -\frac{1}{16}\frac{\sigma^2}{\mu}\Delta + O(\Delta^2) \, .
    \label{eq:gamma_2_neg}
\end{equation}

For the symmetric-distributed $K$:
\begin{equation}
    \gamma_2 = \frac{1}{4}(2\sigma^2)^\frac{1}{3}\Delta^\frac{1}{3} + O(\Delta) \, .
    \label{eq:gamma_2_sim}
\end{equation}

For $K$ distributed in the positive domain:
\begin{equation}
    \gamma_2 = \frac{1}{8}\frac{\sigma^2}{\mu}\Delta + O(\Delta^2) \, .
    \label{eq:gamma_2_pos}
\end{equation}

It can be noted, that for $K$ distributed in the negative domain
$\gamma_2 \not\to0$ as $\Delta\to0$. Our interpretation of the fact is as follows.
The scalings (\ref{eq:gamma_2_neg}) and (\ref{gamma_neg}) refer to the model where
the solution grows in each renovation interval.
For the vanishing $\Delta$ the noise effects vanish and we obtain the positive growth rate 
associated with the mean curvature, i.e. a non-random contribution to the instability.
The scaling (\ref{eq:gamma_2_pos}) deals with the instability intrinsically
related to the noise in the curvature. When  $\Delta \to 0$ the noise
contribution vanishes, and the growth rate also vanishes.
It is also interesting to note that while (\ref{gamma_pos}) yields that the mean of the oscillating field vanishes, the mean energy grows exponentially according to (\ref{eq:gamma_2_pos}).

\begin{figure}[h]
\centering
\includegraphics[width=0.48\textwidth]{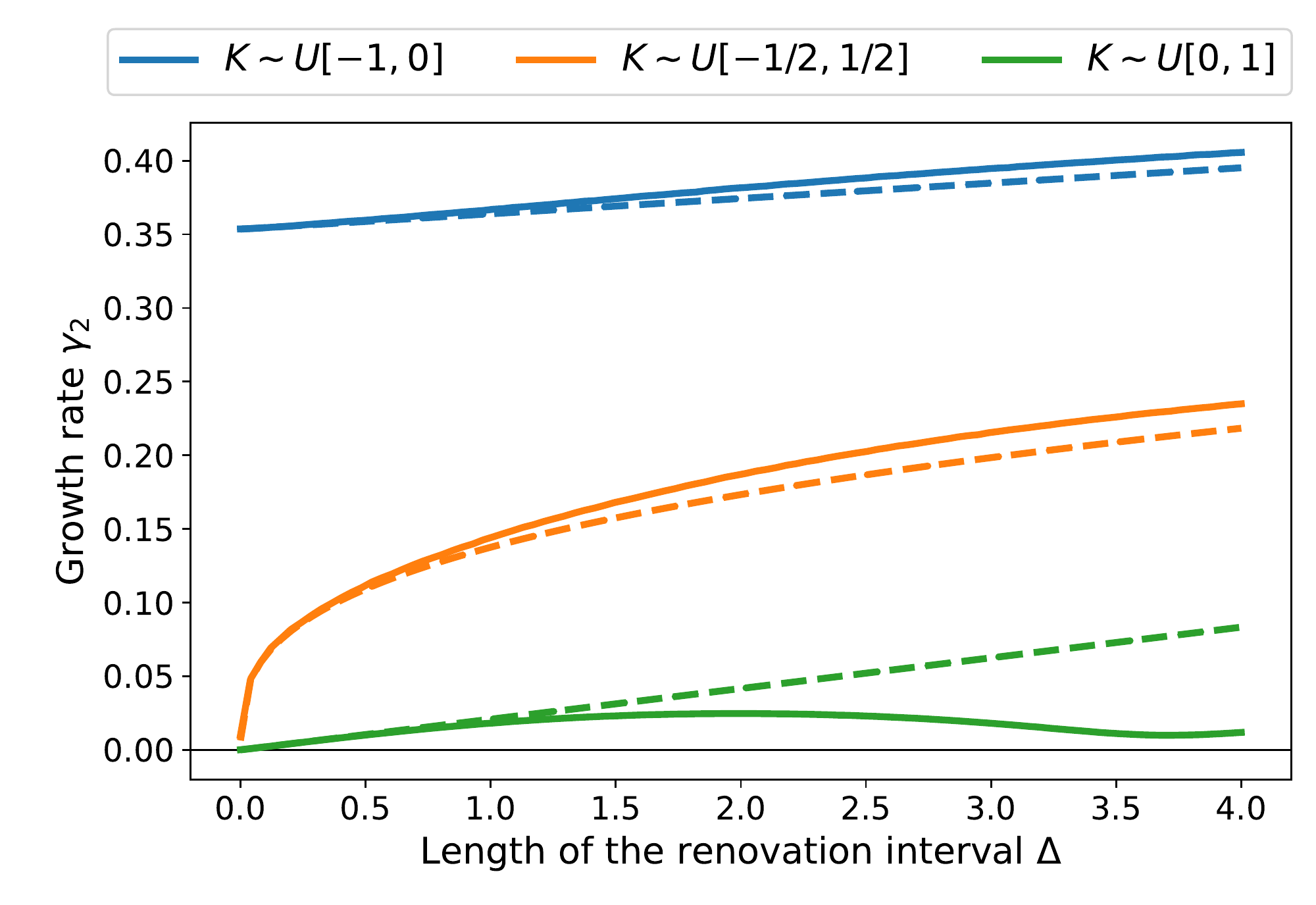}
\caption{Growth rate $\gamma_2$ for different distributions of $K$. Solid lines
show numerical estimates, dashed lines show analytical results}
\label{fig:gamma2}
\end{figure}

Fig.~\ref{fig:gamma2} illustrates the estimates of the growth rate $\gamma_2$
obtained numerically and following (\ref{eq:gamma_2_neg})--(\ref{eq:gamma_2_pos})
for three cases of uniform distribution of~$K$. The numerical approach is similar to those
used in Fig.~\ref{fig:gamma1}. Again, we observe that the analytical 
approximations are in agreement with 
the numerical simulations for small $\Delta$. The agreement for larger $\Delta$ can be improved by
adding terms of the higher orders in series (\ref{eq:gamma_2_neg})--(\ref{eq:gamma_2_pos}), but
they will include higher statistical moments of the random variable $K$. 

\section{Conclusions and discussion}

We obtained that self-excitation with random positive growth rates is more efficient than the competition of
self-excitation and oscillations, which, in turn, is more efficient than the cumulative resonance effect of
random oscillations. This message looks quite natural, but the fact that random oscillations lead to instability
is not so obvious. The latter statement follows from the general mathematical theory \cite{tutubalin1977},
but the quantitative relationship between the three above-mentioned growth rates remained untouched by previous studies. 

Consider the above results in the light of the dynamo problem. There are two realizations
of general dynamo process that are believed to be relevant for, say, solar physics.
First, there is a conventional solar dynamo based on a joint action of differential rotation
and mirror asymmetry of solar convection. The both drivers taken together close the chain of
self-excitation and provide the growing magnetic field. The growth is noisy, however, it becomes
more and more regular if the noise contribution disappears ($\Delta$ becomes small in comparison to the dynamo time scale).
Note that in this paper we considered the limiting case $\Delta \to 0$ and the noise amplitude remained fixed.
Of course, one can also consider the case $\Delta \to 0$, while  the noise amplitude becomes larger as
an appropriate power of $1/\Delta$ to get a non-vanishing growth rate (a short-correlated model).

Another type of dynamo mechanism, possibly related to the solar dynamo, is the so-called small-scale dynamo.
This mechanism contains no specific driving forces. It is based only on random velocity fluctuations and can be
compared to the third case in our analysis, i.e., random positive $K$. If the noisy contribution disappears,
the growth rate becomes negligibly small and the generated magnetic field remains noisy.

Of course, the above messages can be derived from extensive direct numerical MHD modeling, but it is useful to know
that the results obtained from simulations reflect general properties of dynamical systems and can be supported by
examples available for analytical analysis. 

Our analysis is performed within the simple framework proposed in \cite{Z64}, but as far as we know,
the problem is not specific, and analysis of more complicated systems requiring a more bulky algebra is not expected
to lead to more dramatic results (see, e.g., \cite{IS21}). It is also an open question how
the more complicated forms of the random process can influence the form
and stability of the amplification of the mean field and mean energy.
We hope to address these questions in future research.


\begin{acknowledgments}
EI thanks Lomonosov-2 supercomputer center at MSU for computing resources.
DS thanks support by BASIS fund number 21-1-1-4-1.

\end{acknowledgments}

\bibliography{literature}

\end{document}